\documentclass[
reprint,  
amsmath,amssymb,
aps,
prxquantum,
superscriptaddress,
]{revtex4-2}
\usepackage[utf8]{inputenc}
\usepackage[version=3]{mhchem}
\usepackage{amsmath}
\usepackage{color}
\usepackage{titlesec}
\usepackage{soul}
\usepackage{braket}
\usepackage{xr-hyper}
\usepackage{hyperref}
\usepackage{graphicx}
\usepackage{dcolumn}
\usepackage{bm}
\usepackage{tikz}
\usepackage{ragged2e}
\usepackage{kantlipsum}
\usepackage{setspace}
\usepackage{upgreek}
\usepackage{capt-of}
\usepackage{float}
\usepackage{cancel}
\usepackage{svg}
\usepackage[normalem]{ulem}
\allowdisplaybreaks 
\usetikzlibrary{matrix}
\usepackage{eqparbox}
\newbox\matrixcellbox
\tikzset{center align per column/.style={nodes={execute at begin
			node={\setbox\matrixcellbox=\hbox\bgroup},
			execute at end
			node={\egroup\eqmakebox[\tikzmatrixname\the\pgfmatrixcurrentcolumn][c]{\copy\matrixcellbox}}}},
}

\titleformat{\section}{\centering\normalfont\bfseries}{\thesection}{1em}{}
\titlespacing{\section}{0pt}{*2}{*1}
\titlespacing{\subsection}{0pt}{*2}{*1}

\usepackage{xcolor}
\usepackage[draft,inline,nomargin,index]{fixme}

\newcommand{\rb}{r_\mathrm{b}}
\newcommand{\OD}{\mathrm{OD}}

\newcommand {\rmi}{{i}}

\newcommand {\rmd}{{d}}

\newcommand{\Oc}{\Omega}

\newcommand{\DeltaP}{\Delta}

\newcommand{\xo}{x_\mathrm{o}}
\newcommand{\zo}{z_\mathrm{o}}

\newcommand{\len}{L}

\begin{document}
	\onecolumngrid
	\twocolumngrid
	\title{Spatial correlations of photons interacting via transverse Rydberg blockade}
	
	\author{Bankim Chandra Das}
	\email{bankim.das@weizmann.ac.il}
	\affiliation{Department of Physics of Complex Systems, Weizmann Institute of Science, Rehovot 7610001, Israel}
	
	\author{Daniil Svirskiy}
	\affiliation{Institut f\"{u}r Angewandte Physik, University of Bonn, Wegelerstr. 8, 53115 Bonn, Germany}
	\author{Matthias Metternich}
	\affiliation{Institut f\"{u}r Angewandte Physik, University of Bonn, Wegelerstr. 8, 53115 Bonn, Germany}
	\author{Wolfgang Alt}
	\affiliation{Institut f\"{u}r Angewandte Physik, University of Bonn, Wegelerstr. 8, 53115 Bonn, Germany}
	\author{Sebastian Hofferberth}
	\affiliation{Institut f\"{u}r Angewandte Physik, University of Bonn, Wegelerstr. 8, 53115 Bonn, Germany}
	\author{Ofer Firstenberg}
	\affiliation{Department of Physics of Complex Systems, Weizmann Institute of Science, Rehovot 7610001, Israel}

	\date{\today}
	
	\begin{abstract}
		We develop a theory of the transverse spatial dynamics of two interacting photons in a Rydberg nonlinear medium. Extending the well-studied one-dimensional case, we explore both longitudinal and transverse correlations of photons propagating as Rydberg polaritons. We identify distinct behaviors and scaling laws for these correlations, arising from fundamentally different mechanisms in the two directions: diffraction for transverse correlations and diffusion for longitudinal correlations. We develop a model incorporating a Gaussian optical beam and an inhomogeneous atomic density distribution, from which we derive quantitative predictions for correlation functions in both directions. We further show that the propagation equations can be reduced to a single Schrödinger-like equation, allowing approximate solutions that are supported by full numerical results. Our findings indicate that the transverse correlation length is determined primarily by the blockade radius, whereas the longitudinal correlation length is limited by bandwidth. These results establish transverse Rydberg blockade as a distinct and measurable correlation mechanism and show that spatial photon correlations provide a direct means of measuring the blockade radius.

	\end{abstract}
	
	\maketitle
	
	\clearpage
	
	\section{Introduction}
	Achieving strong optical nonlinearities at the level of individual photons is a central goal of quantum optics, motivated by photonic quantum information processing, the engineering of nonclassical states of light, and the exploration of many-body physics with photons \cite{carusotto2013quantum,roy2017colloquium,sheremet2023,chang2014quantum}.
	Effective photon-photon interactions have been realized on several platforms, including atoms and quantum dots coupled to optical cavities and nanophotonic structures \cite{Reiserer2015,birnbaum2005photon,Deutsch1997,Lodahl2015,faraon2008coherent,le2022dynamical,thompson2013coupling,stute2012tunable}, superconducting circuits \cite{hoi2013giant}, and free-space atomic ensembles \cite{chang2014quantum}.
	Among these, a quantum nonlinear optical medium with Rydberg polaritons has emerged
	as one of the leading platforms for studying strong
	photon-photon interactions \cite{Firstenberg2016NonlinearQuantum}.
	This platform has enabled photon blockade and single-photon transistors \cite{Peyronel2012QuantumNonlinear,Gorniaczyk2014SinglePhoton,Rempe2014SinglePhotSwitchPRL,Ornelas2021}, conditional phase shifts and photon-photon gates \cite{Tiarks2016SciAdv,Thompson2017Nature,Tiarks2019NatPhys,Stolz2022PRX}, few-photon bound states \cite{firstenberg2013attractive,cantu2020repulsive}, photon subtraction \cite{murray2018photon,stiesdal2021controlled}, and proposals and observations of many-body phenomena ranging from crystallization to topological defects \cite{chang2008crystallization,Otterbach2013WignerCrystallization,jia2018strongly,clark2020observation,drori2023quantum,das2025multiband}. The wide range of experimental achievements is supported by a rigorous theoretical framework that quantitatively describes quantum correlations arising from the propagation and the interaction of photons \cite{Gorshkov2011PhotonPhoton,Moos2015ManyBody,bienias2014scattering,Gullans2016EffectiveField,Jachymski2016,Grankin2018,Kalinowski2024,CanevaChang2015NJP,Chen2020,Chang2022}.

	To date, however, nearly all free-space experiments and theories
	treat this interaction in one dimension, where the width of the optical beam of interacting photons is smaller than the interaction length set by the Rydberg blockade radius $\rb$. While this geometry yields rich correlation phenomena, all measurable correlations have so far been limited to the temporal domain. 
	But the blockade is an intrinsically three-dimensional phenomenon. The correlations in the transverse direction, which imprint on the spatial domain, remain unexplored at the quantum level. 
	Existing multi-dimensional treatments rely on mean-field or semiclassical descriptions of the nonlinearity \cite{sevinccli2011nonlocal,PhysRevResearch.6.033065, PhysRevLett.115.083602}, while the interaction of transversely separated photons has been demonstrated only for stored, stationary excitations \cite{busche2017contactless}, not for the propagating two-photon wavefunction and its continuous spatial correlations.
	
	Accessing the transverse direction is not merely a higher-dimensional reformulation of the 1D problem. The transverse spatial correlations constitute a natural resource for higher-dimensional photonic quantum states, offering additional degrees of freedom for quantum communication and computation \cite{yang2025efficientmultiplexedquantummemory,PhysRevLett.114.050502,Wang2024UltrahighFidelitySpatialMode}, with the blockade providing a direct, interaction-based route to their generation. It also opens the spatial domain to phenomena so far identified only in the time domain, such as quantum vortices and other topological defects of the two-photon wavefunction \cite{drori2023quantum,das2025multiband}.

	In this work, we develop a theoretical model of two-photon transverse interactions in a two-dimensional Rydberg medium, retaining the spatial extent of both the probe beam and the atomic cloud. That is, we consider one transverse dimension ($z$) and one longitudinal dimension ($x$). While this formulation can be used for both dissipative \cite{Peyronel2012QuantumNonlinear} and dispersive \cite{firstenberg2013attractive} regimes of photon interactions, in this work, we consider only the dissipative regime for resonant probe and control fields in a three-level system under conditions of electromagnetically induced transparency (EIT).

	We first compute the stationary two-photon wavefunction of the full nine-component model for a two-dimensional Gaussian cloud with a finite beam waist, and then propagate it to the experimentally accessible second-order correlation functions in both the temporal and spatial domains. 
	While the full numerical model provides details of the wavefunction and the correlation structure, interpreting the description is cumbersome. We further reduce the full model to a single effective Schr\"odinger-like equation in two dimensions whose closed-form structure identifies the blockade mechanism behind each dimension. The kinetic energy coefficient of the relative coordinate is real along the propagation axis, which introduces a diffusion set by the atomic response and the EIT bandwidth. But, in the transverse plane, the kinetic energy coefficient is imaginary, resulting in photonic paraxial diffraction. Furthermore, from this equation we find distinct differences in the correlation length, \textit{e.g.}, in the scaling with optical depth(OD) 
	for the spatial and temporal correlations. 
	These results establish the transverse Rydberg blockade as a distinct correlation mechanism, with spatial photon correlations as its new, measurable observable.

	\begin{figure}
		\includegraphics[width=\linewidth]{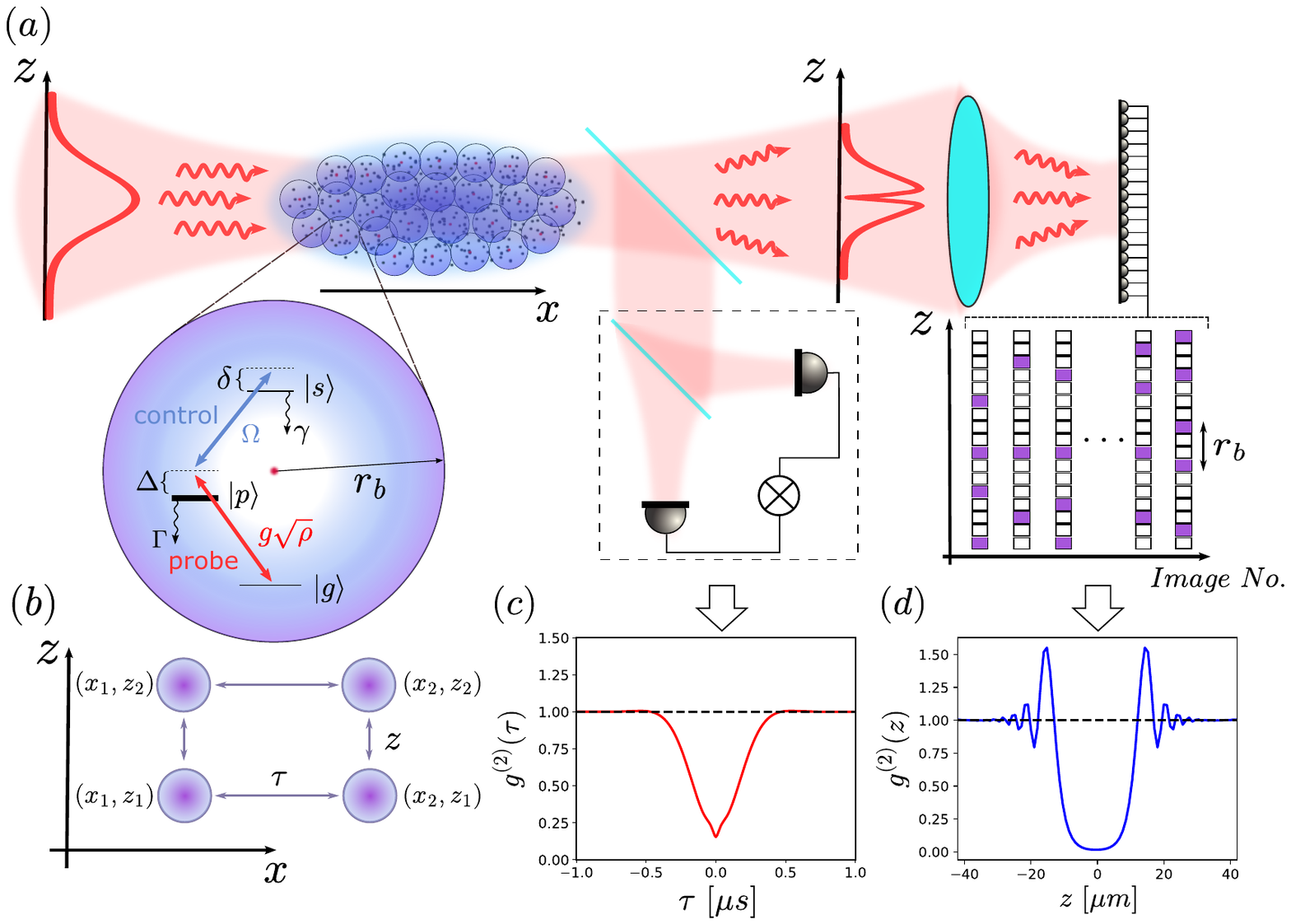}
		\caption{\label{fig:setup}(a) An ultracold large atomic ensemble is excited via two-photon transition to a Rydberg state (the control beam is not shown). We consider a ladder scheme with ground state $\ket{g}$, intermediate state $\ket{p}$ (decay rate $\Gamma$, one-photon detuning $\DeltaP$), and Rydberg state $\ket{s}$ (decoherence rate $\gamma$, two-photon detuning $\delta$); the probe couples $\ket{g}\!\leftrightarrow\!\ket{p}$ with collective strength $g\sqrt{\rho}$, and the control field couples $\ket{p}\!\leftrightarrow\!\ket{s}$ with Rabi frequency $\Oc$. The probing beam waist $w_0$ is chosen such that the blockade radius $\rb$ is smaller than the beam waist $w_0/\rb  > 1$. This condition breaks a typical 1D approximation. We investigate both the temporal correlations (c), which can be measured by a HBT setup, as well as the transverse correlations (d), measured using an array of single-photon avalanche photodiodes or other spatially-resolving single-photon sensitive device. (b) The coordinate system used in the paper has $x$ and $z$ as the relative coordinates in the longitudinal and transverse directions, respectively.} 
	\end{figure}

	\section{Physical system}
	
	We consider an ultracold cloud of three-level atoms that hosts Rydberg polaritons under EIT conditions. 
	The ground $\ket{g}$, excited $\ket{p}$, and Rydberg $\ket{s}$ states are coupled via the ladder-type scheme [see Fig.~\ref{fig:setup}(a)]: \[\ket{g} \overset{g \sqrt{\rho}}{\underset{\Delta,\Gamma}\rightleftarrows} \ket{p}\overset{\Omega}{\underset{\delta,\gamma}\rightleftarrows}\ket{s},\] forming Rydberg polaritons.
	Since we are interested in studying the transverse component of the Rydberg interaction, we consider a probe beam that has a transverse spatial extent, along with a 2D Gaussian cloud $\rho(x,z)=\rho_0 \exp[-\dfrac{x^2}{2\sigma_x^2}-\dfrac{z^2}{2\sigma_z^2}]$ where $x$ is the propagation direction and $z$ is the transverse direction. 
	$L_x=\sqrt{2\pi}\sigma_x, L_z=\sqrt{2\pi}\sigma_z$ are the longitudinal and transverse lengths of the medium, respectively.
	Furthermore, for the Rydberg interaction to act on the photons in the transverse direction, we require  $\sigma_z> \rb$,  where $ \rb = \sqrt[6]{C_6/2\hbar \Gamma_\mathrm{EIT}}$ is the blockade radius, $\Gamma_\mathrm{EIT}=\Omega^2/\Gamma$ and $C_6$ is the van der Waals interaction coefficient \cite{Firstenberg2016NonlinearQuantum}. 
	The input electric field has a transverse amplitude profile  $E(z)=\exp[-z^2/w_0^2]$ with a beam waist $w_0$.
	The longitudinal distance between photons in the medium becomes their temporal separation once they are outside, and thus can be measured by a typical temporal correlation setup.
	In Fig.~\ref{fig:setup}(c), we show a typical case for dissipative interaction \cite{Peyronel2012QuantumNonlinear}. 
	In contrast, spatially resolved detection of outgoing single photons can reveal transverse correlations. For the dissipative case, one can expect that the outgoing photons will exhibit a similar antibunching behavior in space, which means that the minimal distance between two detector `clicks' should not be smaller than the characteristic distance $\rb$, as sketched in Fig.~\ref{fig:setup}(d).
	
	\begin{figure}
		\includegraphics[width=\linewidth]{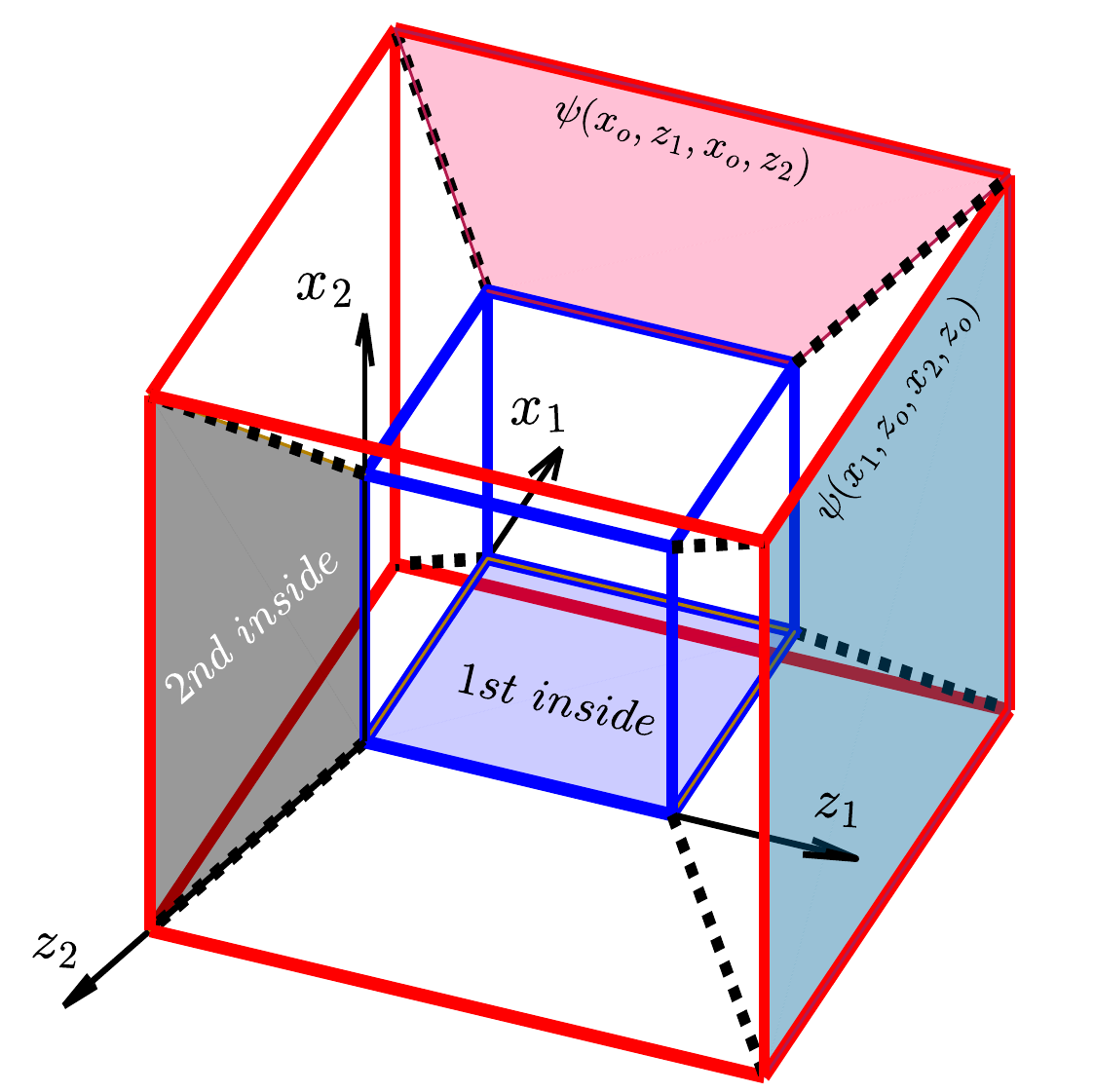}
		\caption{Schematic visualization of the four-dimensional configuration space of the two-photon 
			wavefunction $\psi(x_1,z_1, x_2, z_2)$, represented as a 
			tesseract by two nested cubes (inner blue and outer red). The coordinate pairs $(x_1, z_1)$ and $(x_2, z_2)$ specify the positions of photons 1 and 2, respectively. The coordinates $x_1$, $x_2$, and $z_1$ form the three orthogonal axes of each cube, while the fourth coordinate, $z_2$, is represented by the connections between the cubes. The gray- and blue-shaded `input' planes correspond to initially uncorrelated photons: on the blue (gray) plane, photon 1 (2) enters the medium while photon 2 (1) has not yet entered.    
			Propagating through the atomic cloud 
			generates two-photon correlations through the Rydberg blockade interaction. The longitudinal $\psi(x_1,\zo,x_2,\zo)$ and transverse $\psi(\xo,z_1,\xo,z_2)$ correlations are located in the 'output' planes -- shaded light blue and light red, respectively.}
		\label{fig:tesseract}
	\end{figure}
	
	\section{Numerical methods}\label{Numerics}
	\subsection{Two-photon propagation equations}
	
	\begin{figure}[!t]
		\includegraphics[width=\linewidth]{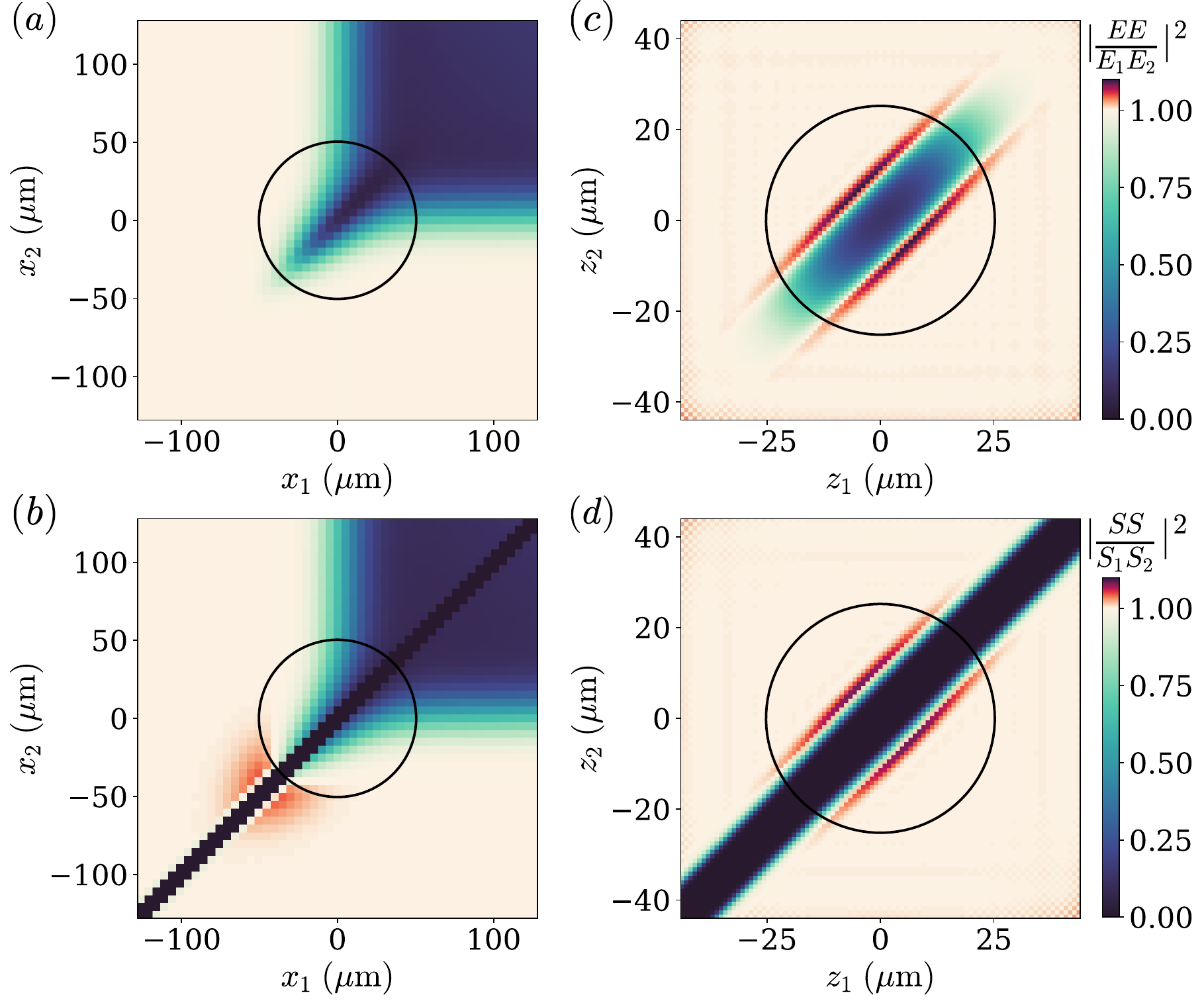}
		
		\caption{Numerically calculated normalized stationary two-photon wavefunction $\psi(x_1,z_1,x_2,z_2)$ for a 2D Gaussian atomic cloud $\rho(x,z)$, indicated by the black circle, showing the Rydberg blockade in both the longitudinal and transverse planes. Each panel shows the two-particle amplitude normalized to the product $i_1j_2$ of the corresponding single-photon amplitudes, defined in Sec.~\ref {Numerics}.
			(a) Two-photon amplitude $\left|\dfrac{EE}{E_1E_2}\right|^2$ in the longitudinal $(x_1,x_2)$-plane at the transverse center of the cloud, $z_1=z_2=0$. The blockade appears as a suppression along the diagonal, whose width grows with propagation and is set by the EIT bandwidth.
			(b) Two-Rydberg amplitude $\left|\dfrac{SS}{S_1S_2}\right|^2$ in the same plane. The dark diagonal confirms that two simultaneous Rydberg excitations are forbidden within the blockade radius, with a sharp, propagation-independent width set by $\rb$.
			(c) Two-photon amplitude $\left|\dfrac{EE}{E_1E_2}\right|^2$ in the transverse $(z_1,z_2)$-plane at the longitudinal center of the cloud, $x_1=x_2=0$, revealing the blockade in the transverse direction. Unlike the longitudinal case, the width here is governed by $\rb$ rather than by the EIT bandwidth.
			(d) Two-Rydberg amplitude $\left|\dfrac{SS}{S_1S_2}\right|^2$ in the same plane. The forbidden region has a fixed width $\sim 2\rb$, in contrast to the broadening of the two-photon amplitude in (a) and (c), which reflects propagation through the medium. Simulation parameters are given in Sec.~\ref{numerical_params}.}
	\label{fig:two_photon_amplitude}
\end{figure}

The  single-polariton propagation through the medium is governed by $H(x,z)\psi=\rmi \partial_t\psi$, where 

\begin{equation}\label{eq:H}
	H(x,z)= 	
	{-}\begin{pmatrix}
		ic\partial_x +\dfrac{c}{2k}\partial^2_z&  \sqrt{\rho(x,z)}g & 0\\	
		\sqrt{\rho(x,z)}g   & \Delta+\rmi\Gamma &  \Omega \\
		0& \Omega & \delta+\rmi\gamma\\
	\end{pmatrix}
\end{equation}
is the effective Hamiltonian for the photon and atoms. The single polariton wavefunction $\psi(x,z)=[E(x,z),P(x,z),S(x,z)]^T$ consists of amplitudes $E(x,z)$ for a single photon (all atoms in the ground state $\ket{g}$), $P(x,z)$ for one atom in the intermediate state $\ket{p}$, and $S(x,z)$ for one atom in the Rydberg state $\ket{s}$.

The equation of motion for the one-photon amplitudes derived from Heisenberg equation \cite{Peyronel2012QuantumNonlinear,Gorshkov2011PhotonPhoton,Moos2015ManyBody} is in the 2D case:
\begin{align}\label{eq:one-photon}
		\dfrac{\partial E}{\partial  t} &= -c \dfrac{\partial E}{\partial x}+ i\dfrac{c}{2k}\nabla^2_z E+ ig\sqrt{\rho(x,z)} P\:\nonumber,\\
		\dfrac{\partial P}{\partial t} &= - (\Gamma- i \Delta)P+ i  g\sqrt{\rho(x,z)} E+ i \Oc S\:\nonumber,\\
		\dfrac{\partial S}{\partial t} &= - (\gamma- i \delta)S+ i  \Oc P\:.
	\end{align}
Here, the first spatial derivative in $x$ of the electric field of the photon represents that the photon propagates in the $x$ direction, whereas the second derivative in the $z$ direction shows the paraxial diffraction component of the input beam.

The Hamiltonian and equation of motion can be straightforwardly generalized to two polaritons, described by a multi-component wavefunction. The two-polariton Hamiltonian can be written as $\mathcal{H}(x_1,z_1,x_2,z_2)=H(x_1,z_1)\otimes I+I\otimes H(x_2,z_2)+\mathcal V$.
Here, $(x_1,z_1)$ are the spatial coordinates for the first photon, and $(x_2,z_2)$ are the spatial coordinates for the second photon [Fig.~\ref{fig:setup}(b)]. The interaction potential $\mathcal V$ acting on the two-Rydberg component $SS$ originates from van der Waals interactions between Rydberg atoms and is given by  
\vspace{-.2cm}
\begin{equation}\label{eq:V}
	\mathcal{V}= \dfrac{1}{2}\!\int\! \rmd x\, \rmd z\, \rmd x'\rmd z'\, V_\mathrm{ss}\,
	\hat{S}^\dagger(x,z)\hat{S}^\dagger(x',z')\hat{S}(x',z')\hat{S}(x,z),
\end{equation}
where $\hat{S}(x,z)$ is the excitation operator to the Rydberg state, and $V_\mathrm{ss}(x_1,z_1,x_2,z_2)= \dfrac{(C_6/\hbar)}{\left[|x_1-x_2|^2+|z_1-z_2|^2\right]^3}$ is the Rydberg interaction potential.

\begin{figure*}
	\includegraphics[width=\textwidth]{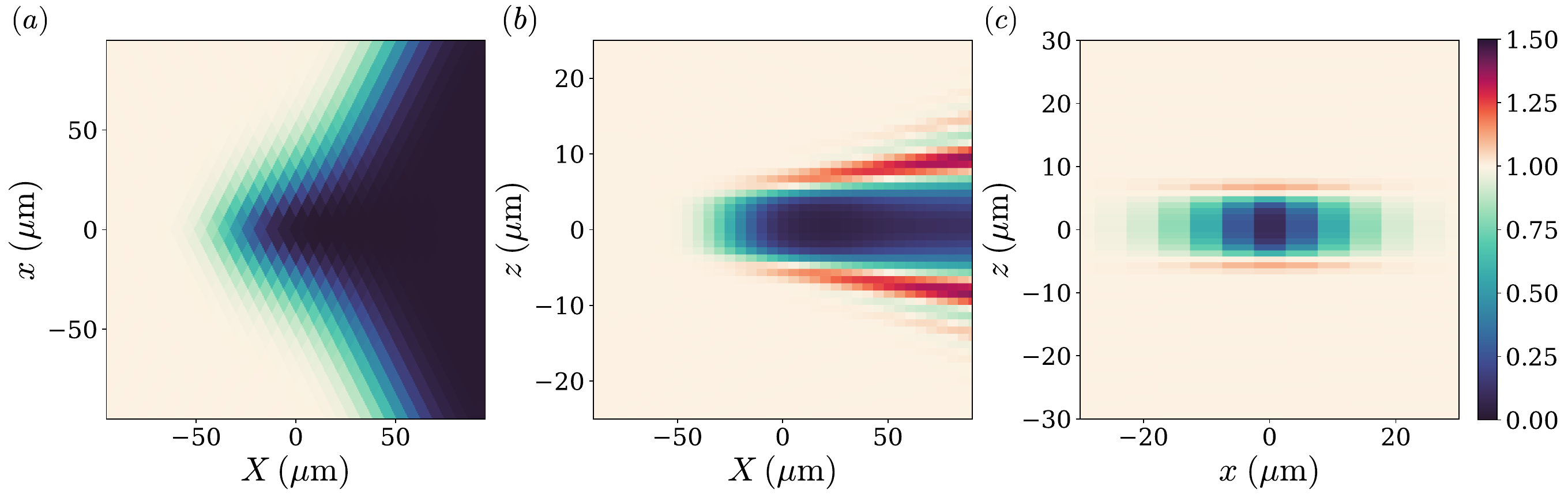}
	\caption{Normalized two-photon wavefunction amplitude $|EE|^2$ plotted in center-of-mass ($X$) and relative coordinates ($x,z$).  Normalized two- photon wavefunction $\left|\dfrac{EE}{E_1E_2}\right|^2$  
		(a) in the $(X,x)$-plane, showing the longitudinal blockade at the center of the cloud. The blockade width grows as $\sqrt{X}$ with propagation, characteristic of the EIT bandwidth-limited response in the longitudinal direction. 
		(b) in the $(X,z)$-plane, showing the transverse blockade at the center of the cloud. Here, the blockade width is set by the Rydberg blockade radius $\sim 2\rb$, independent of the EIT bandwidth. The slow broadening with $X$ reflects the paraxial diffraction of the beam rather than the EIT response. The blockaded region acts as an obstacle, diffracting the beam. The intensity removed from the core accumulates on either side, appearing as bunching and oscillatory fringes.
		(c) in the relative coordinate $(x,z)$-plane, showing the blockade simultaneously in both directions. The blockade region forms an ellipse, with the longitudinal width exceeding the transverse width, reflecting the distinct length scales set by the EIT bandwidth and the transverse blockade diffraction, respectively. The parameters for the simulation are mentioned in section \ref{numerical_params}.
	}
	\label{fig:two_photon_amplitude_trans}
\end{figure*}

We shall consider an atomic cloud with a Gaussian density profile $\rho(x,z)$ centered at the origin of the coordinate system. For the numerical calculations, we take a bounding box of dimensions $2\xo\times 2\zo$, such that the range $x\in [-\xo,\xo]$ and $z\in [-\zo,\zo]$ fully contains the cloud (usually $\xo=5\sigma_x$ and $\zo=3.7\sigma_z$). 
As the two-photon wavefunction at time $t$ depends on four spatial coordinates, $x_1$, $x_2$, $z_1$, $ z_2$, the structure of the problem can be represented as a four-dimensional hypercube, \textit{i.e.}, a tesseract, as illustrated in
Fig.~\ref{fig:tesseract}. 
To simplify the visualization, only the input and output correlation planes in the longitudinal and transverse directions are shown. 

We denote $\psi(t,x_1,z_1,x_2,z_2) \equiv \psi$, which consists of nine components \cite{Peyronel2012QuantumNonlinear, Gorshkov2011PhotonPhoton,drori2023quantum}.
The equations of motion for the two-photon amplitudes then read: 
\begin{widetext}
	\begin{align}\label{eqs:full2}
		\dfrac{\partial EE}{\partial t}&= -c \left(\dfrac{\partial}{\partial x_1}+\dfrac{\partial}{\partial x_2}\right)EE +i\dfrac{c}{2k}\left(\nabla^2_{z_1}+\nabla^2_{z_2}\right) EE+ i g \left[\sqrt{\rho(x_1,z_1)}PE+\sqrt{\rho(x_2,z_2)}EP\right]\:,\\\nonumber
		\dfrac{\partial EP}{\partial t}&=- c \dfrac{\partial}{\partial x_1}EP+i\dfrac{c}{2k}\nabla^2_{z_1} EP-(\Gamma-i\DeltaP)EP + ig \left[\sqrt{\rho(x_1,z_1)}PP+\sqrt{\rho(x_2,z_2)}EE\right]+ i \Oc ES\:,\\\nonumber
		\dfrac{\partial ES}{\partial t}&=-c \dfrac{\partial}{\partial x_1}ES+i\dfrac{c}{2k}\nabla^2_{z_1} ES-(\gamma-i\delta)ES + i g\sqrt{\rho(x_1,z_1)}PS+ i \Oc EP\:,\\\nonumber
		\dfrac{\partial PE}{\partial t}&=- c \dfrac{\partial}{\partial x_2}PE+i\dfrac{c}{2k}\nabla^2_{z_2} PE-(\Gamma-i\DeltaP)PE+ ig\left[\sqrt{\rho(x_2,z_2)}PP+\sqrt{\rho(x_1,z_1)}EE\right]+ i \Oc SE\:,\\\nonumber
		\dfrac{\partial PP}{\partial t}&=-2(\Gamma-i\DeltaP)PP + i g\left[\sqrt{\rho(x_1,z_1)}EP+\sqrt{\rho(x_2,z_2)}PE\right]+ i \Oc\left[PS+ SP\right]\:,\\\nonumber
		\dfrac{\partial PS}{\partial t}&=-(\Gamma-i\DeltaP)PS-(\gamma-i\delta)PS + i g \sqrt{\rho(x_1,z_1)}ES+i \Oc\left[SS+PP\right]\:,\\\nonumber
		\dfrac{\partial SE}{\partial t}&=- c\dfrac{\partial}{\partial x_2}SE+i\dfrac{c}{2k}\nabla^2_{z_2} SE-(\gamma-i\delta)SE + ig\sqrt{\rho(x_2,z_2)}SP+i \Oc PE\:,\\\nonumber
		\dfrac{\partial SP}{\partial t}&=-(\Gamma-i\DeltaP)SP-(\gamma-i\delta)SP + i g\sqrt{\rho(x_2,z_2)}SE+i \Oc\left[SS+PP\right]\:,\\\nonumber
		\dfrac{\partial SS}{\partial t}&=-2(\gamma-i\delta)SS +i \Oc\left[PS+SP\right]-i V_{ss}(x_1,x_2,z_1,z_2)SS\:,\nonumber
\end{align}
\end{widetext}
These nine equations can be solved to obtain the steady-state two-photon amplitude $EE(x_1,z_1,x_2,z_2)$ inside the medium.

\subsection{Numerical solution for a Gaussian density}
To obtain the two-photon steady-state solution, we numerically evolve Eqs.~(\ref{eqs:full2}) until convergence at a sufficiently large time $t$.  
We first solve Eq.~(\ref{eq:one-photon}) for the single-polariton steady-state wavefunction $\psi_i(x,z)$ with $i=\{E,P,S\}$, which provides the stationary boundary conditions for the nine-component Eq.~(\ref{eqs:full2}) \cite{Peyronel2012QuantumNonlinear,firstenberg2013attractive,drori2023quantum,das2025multiband},
\begin{gather}
\begin{aligned}
	\psi_{ij}(x_1=-\xo,z_1,x_2,z_2,t)&=\psi_{i}(x_1=-\xo,z_1)\psi_{j}(x_2,z_2),\\
	\psi_{ij}(x_1,z_1,x_2=-\xo,z_2,t)&=\psi_{j}(x_2=-\xo,z_2)\psi_{i}(x_1,z_1).\\
\end{aligned}
\end{gather}
These $\psi_{ij}$ with $i,j=\{E,P,S\}$ reside in the gray and blue shaded planes in Fig.~\ref{fig:tesseract}. In the following we abbreviate the single-photon amplitudes of photons $1$ and $2$ as $i_1\equiv\psi_i(x_1,z_1)$ and $j_2\equiv\psi_j(x_2,z_2)$, so that the products above read $i_1j_2$.
We then solve Eqs.~(\ref{eqs:full2}) for the two-photon steady-state amplitude $\psi_{ij}(x_1,z_1,x_2,z_2)$ with the initial condition $\psi(x_1,z_1,x_2,z_2,t=0)=0$.
The interaction potential $V_{ss}$ produces a blockade in both the longitudinal and transverse directions, forbidding two simultaneous Rydberg excitations ($SS=0$) when the photons are close, $x_1\approx x_2$ or $z_1\approx z_2$.

Figure \ref{fig:two_photon_amplitude} shows the normalized steady-state amplitude of the full numerical solution for the two-photon ($EE$) and two-Rydberg ($SS$) components, in both the longitudinal $(x_1,x_2)$ and transverse $(z_1,z_2)$-planes at the center of the cloud. 
As expected, the two photons block each other when their relative distance falls below $\sim\rb$.
The two-Rydberg component [Figs.~\ref{fig:two_photon_amplitude}(b,d)] shows forbidden regions along the diagonal of the longitudinal and transverse planes, and the two-photon amplitude [Figs.~\ref{fig:two_photon_amplitude}(a,c)] shows the same behavior. 

Interestingly, the blockade broadens through distinct mechanisms in the longitudinal and transverse directions. 
In Eq.~\eqref{eqs:full2}, the two-photon amplitude $EE$ contains a first-order derivative along the propagation direction $x$ and a second-order derivative with a complex coefficient along the transverse direction $z$. In the effective two-photon description of Sec.~\ref{effective Schrodinger}, the former combines with the finite EIT bandwidth \cite{Peyronel2012QuantumNonlinear} to give a diffusion in the relative coordinate, whereas the latter produces no diffusion and instead imprints a phase and a paraxial diffraction that only slightly modifies the blockade profile, whose width remains set mainly by the Rydberg blockade diameter $\sim2\rb$.

The two length scales are best visualized by plotting the amplitudes in center-of-mass and relative coordinates. We define $X=(x_1+x_2)/2$ and $x=x_1-x_2$ \cite{Peyronel2012QuantumNonlinear} as the center-of-mass and relative coordinates in the longitudinal direction, and $Z=(z_1+z_2)/2$ and $z=z_1-z_2$ as their transverse counterparts.
Here $X$ is the effective propagation direction of the photons, i.e., the diagonal of the $(x_1,x_2)$-plane. Figures~\ref{fig:two_photon_amplitude_trans}(a,b) show the normalized amplitude $EE$ in the $(X,x)$ and $(X,z)$-planes.
In Fig.~\ref{fig:two_photon_amplitude_trans}(a), the blockade width grows as $\sqrt{X}$, as is well established for 1D systems \cite{Peyronel2012QuantumNonlinear}.
In Fig.~\ref{fig:two_photon_amplitude_trans}(b), the width is set mainly by the blockade diameter $\approx 2\rb$ and broadens only slowly with propagation.

\begin{figure}
\includegraphics[width=\linewidth]{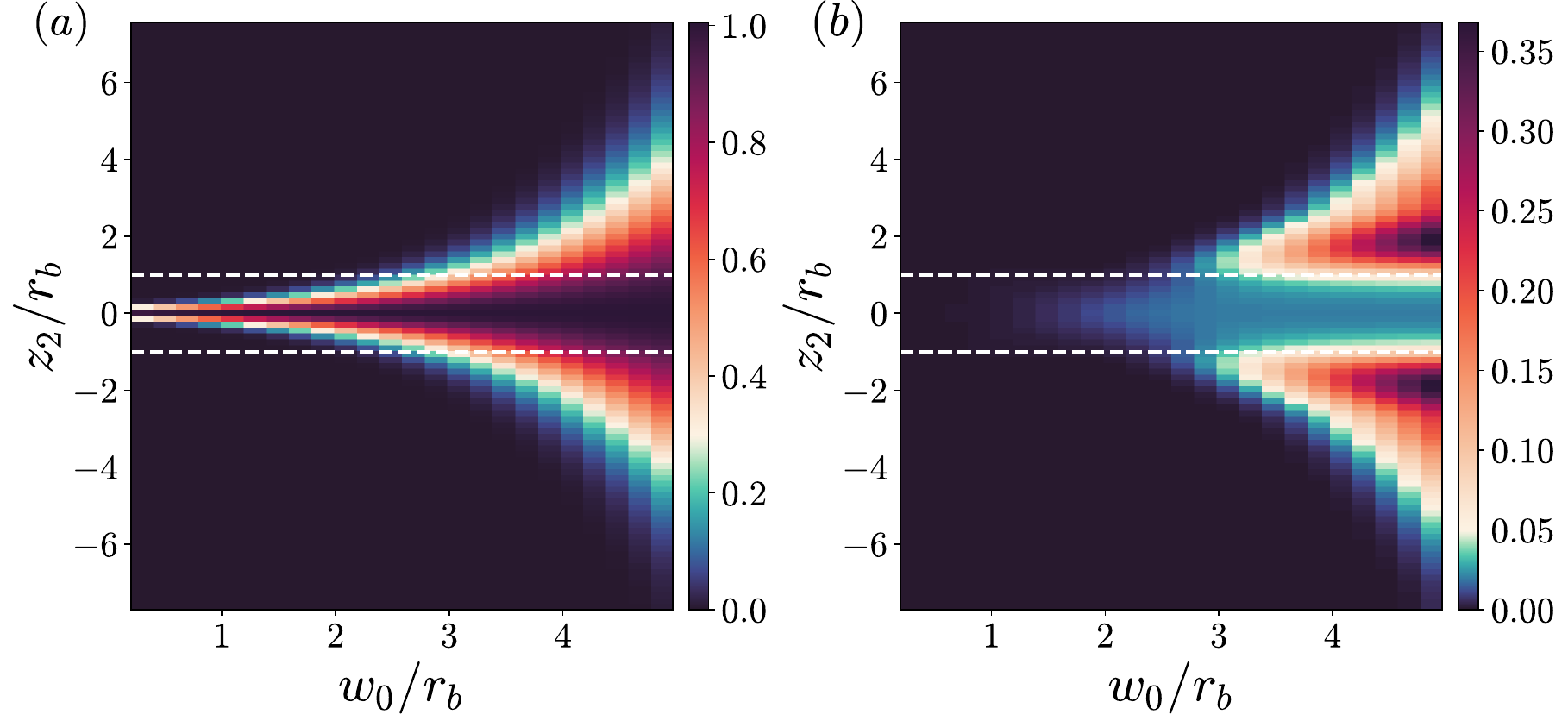}
\caption{Numerically calculated transverse profile of the two-photon amplitude $|EE(\xo,0,x_2,z_2)|^2$, with one photon detected at the exit of the medium ($x_1=\xo$), as a function of normalized transverse position $z_2$ and beam waist $w_0$, at (a) the entrance ($x_2=-\xo$) and (b) the exit ($x_2=\xo$) of the atomic medium. Comparing the two reveals the effect of the Rydberg blockade on the transmitted beam. For $w_0 \gg \rb$, the blockade imprints a central depletion hole of width $2\rb$ on the output profile, confirming that the suppression is blockade-induced. For $w_0 \leq \rb$, the blockade weakens and the output profile is instead broadened by paraxial diffraction, with no clear central hole. The crossover near $w_0 \approx \rb$ marks the transition between the blockade and diffraction-dominated regimes. 
	Simulation parameters are given in Sec.~\ref{numerical_params}.}
\label{w0 variation}
\end{figure}

Finally, Fig.~\ref{fig:two_photon_amplitude_trans}(c) shows the wavefunction in the $(x,z)$-plane at the exit of the medium.
The different length scales in the two directions make the blockaded region elliptical, as expected. 
To clarify the paraxial contribution to the transverse blockade width, we vary the input beam waist $w_0$ relative to the blockade radius $\rb$ [Fig.~\ref{w0 variation}].
To isolate the propagation effect from that of the cloud shape, we here consider a uniform 2D cloud of lengths $L_x$ and $L_z$ in the longitudinal and transverse directions.
Figures~\ref{w0 variation}(a,b) show the transverse profile of the two-photon wavefunction, which at the entrance reproduces the Gaussian probe beam and at the exit carries the imprint of the blockade.
For large $w_0/\rb$, the blockade width is set mainly by the blockade diameter $2\rb$ and the paraxial effect is negligible. For $w_0/\rb \ll 1$, the paraxial effect dominates and the transverse blockade is strongly suppressed.

\section{Effective two-photon Schrödinger equation}\label{effective Schrodinger} 
While the previous section shows that Eqs.~\eqref{eqs:full2} can be solved numerically for realistic experimental conditions, an approximate closed-form equation offers a more instructive account of the longitudinal and transverse correlations imprinted by the Rydberg blockade. Such effective equations are well established in Rydberg polariton systems \cite{Peyronel2012QuantumNonlinear,drori2023quantum,firstenberg2013attractive}.
Following \cite{das2025multiband,drori2023quantum}, we generalize the effective 1D Schr\"odinger equation to 2D via a Dirac-like two-component equation, using the center-of-mass $(X,Z)$ and relative $(x,z)$ coordinates introduced in the previous section.
We further define the symmetric and antisymmetric combinations of the amplitudes:
\begin{align} ES_{\pm}= \dfrac{ES\pm SE}{2},\hspace{-4mm} \quad PS_{\pm}= \dfrac{PS\pm SP}{2},\hspace{-4mm} \quad EP_{\pm}= \dfrac{EP\pm PE}{2}\:.\end{align} 
Next, we eliminate $PP$, $PS_{\pm}$, $EP_{\pm}$, and $SS$ from Eqs.~\eqref{eqs:full2}.
For simplicity, we set $\gamma=0$ and $\DeltaP=0$ and consider a homogeneous cloud of lengths $L=L_x=\sqrt{2\pi}\sigma_x$ and $L_z=\sqrt{2\pi}\sigma_z$. 
The elimination yields the following Dirac-like equation for the amplitudes $ES_{\pm}$:
\begin{multline}\label{eq:dirac}
\dfrac{\partial}{\partial X} 
\begin{pmatrix}
	ES_+\\ES_-
\end{pmatrix}= \\
\begin{pmatrix} \dfrac{\rmi}{k}\dfrac{\partial^2}{\partial z^2}-\dfrac{\OD}{L}V^{(2)}(x,z)& - 2\dfrac{\partial}{\partial x}\\
	- 2\dfrac{\partial}{\partial x} &
	\dfrac{\rmi}{k}\dfrac{\partial^2}{\partial z^2}-\dfrac{\OD}{L}
\end{pmatrix}
\begin{pmatrix}
	ES_+\\ES_-
\end{pmatrix}\:,
\end{multline}
where 
$V^{(2)}$ is the effective Rydberg interaction potential, given by $1/V^{(2)}=1-i\left[\left(\dfrac{x}{\tilde{\rb}}\right)^2+\left(\dfrac{z}{\tilde{\rb}}\right)^2\right]^3$, and $\tilde{\rb}=\rb/\sqrt[6]{2}$ is the modified blockade radius. Here we drop the second derivative with respect to the center-of-mass coordinate $Z=(z_1+z_2)/2$.
Adiabatically eliminating $ES_-$ in Eq.~\eqref{eq:dirac}, we obtain a single-component effective two-photon Schr\"odinger-like equation in the 2D plane:
\begin{align}\label{eq: schrodinger}
\dfrac{\partial}{\partial X} \psi&= \dfrac{4L}{\OD} \dfrac{\partial^2}{\partial x^2} \psi +\dfrac{i}{k}\dfrac{\partial^2}{\partial z^2}\psi - \dfrac{\OD}{L}V^{(2)}(x,z)\psi \:. 
\end{align}
Here $\psi=ES_+(X,x,z)$ is proportional to the two-photon amplitude $EE(X,x,z)$ \cite{firstenberg2013attractive}, so that the normalized $ES_+$ gives the two-photon wavefunction.
Equation~\eqref{eq: schrodinger} resembles its 1D counterpart \cite{Peyronel2012QuantumNonlinear}, but with a 2D effective interaction potential $V^{(2)}(x,z)$ that includes a transverse blockade component, and with an additional paraxial diffraction term $\frac{i}{k}\frac{\partial^2}{\partial z^2}$ carrying an imaginary coefficient.
The depleted region created by the blockade interaction $V^{(2)}(x,z)$ is then further modified by the two mass-like terms in the longitudinal and transverse directions.

The two kinetic terms are therefore anisotropic. The longitudinal coefficient depends on the polariton response, whereas the transverse one is purely photonic.
Even for a symmetric potential, the blockade shape in the $(x,z)$-plane is elliptical and governed by the relative strengths of these two terms.
The longitudinal mass term $(OD/8L)$ acts as a diffusion coefficient along the probe propagation direction, setting the width of the depletion region as
$w_{\parallel}(X) \propto \sqrt{\frac{8L}{\OD}X}$.
In the transverse direction, there is no diffusion. Instead, paraxial diffraction broadens the depletion region asymptotically as
$w_{\bot}(X) \propto \sqrt[8]{\frac{\OD}{kL}X^2}$.
Both widths are converging to a characteristic value of an effective blockade radius $2r_{eff}$.
These scalings agree with the full numerical results of the previous section. Their derivation is given in Appendix~\ref{app:derivation}.\\

\begin{figure}
\includegraphics[width=\linewidth]{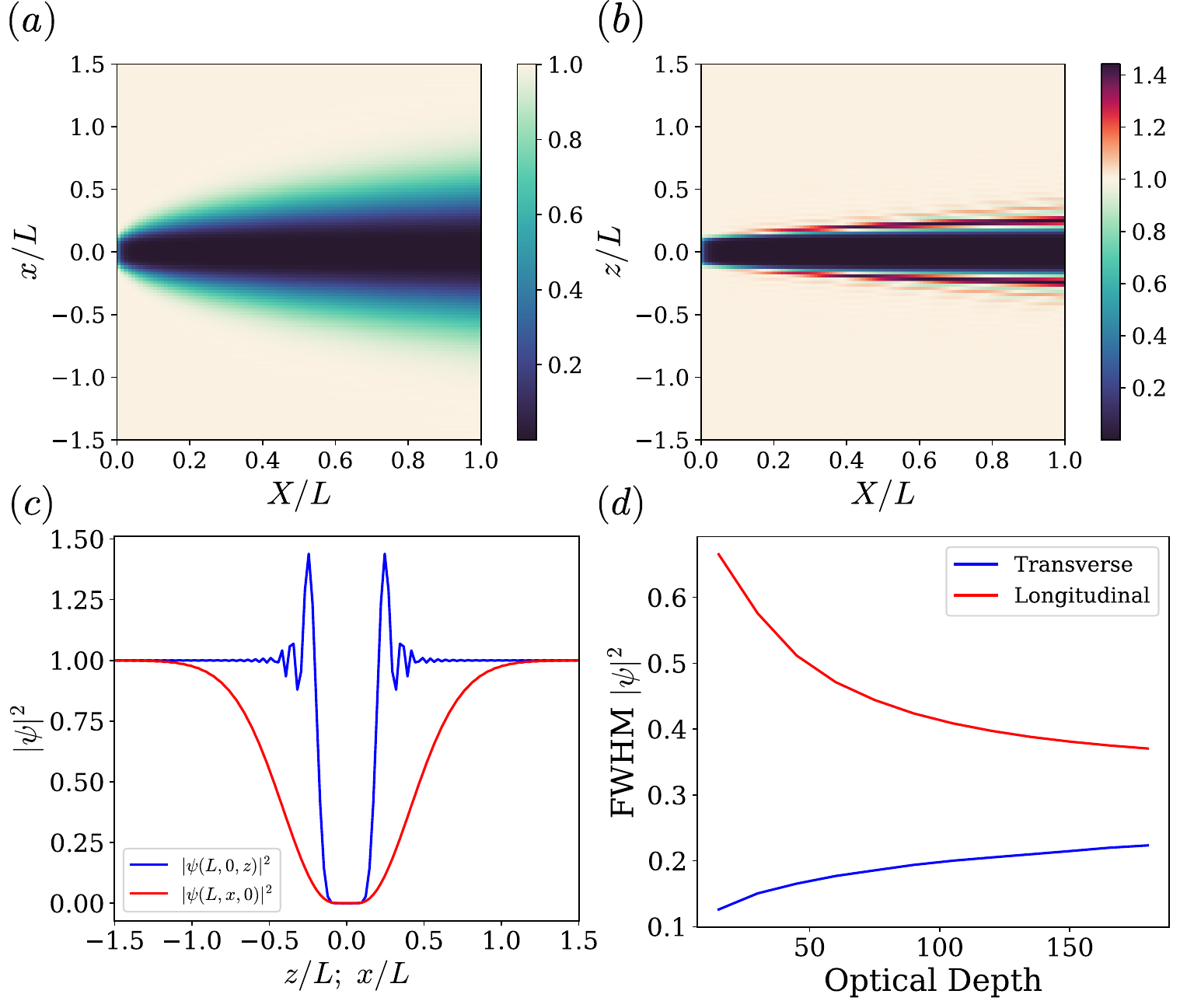}
\caption{Longitudinal and transverse dynamics along the medium with $\OD=60$ and $\rb=7.4$~$\mu$m, obtained from the single 2D Schrödinger equation~\eqref{eq: schrodinger}. (a,b) Normalized two-photon wavefunction in two orthogonal planes, $|\psi(X,x,0)|^2$ (a) and $|\psi(X,0,z)|^2$ (b). The photon behaves as a massive particle in an effective 2D potential, as in the 1D case, but the transverse blockade is now visible owing to the transverse Rydberg interaction.
	(c) $|\psi|^2$ along the longitudinal (red) and transverse (blue) directions at the exit of the medium, $X=L$, showing the diffraction-like behavior.
	(d) Full width at half maximum (FWHM) of $|\psi|^{2}$ as a function of the optical depth of the medium.  
	The longitudinal blockade width decreases with increasing $\OD$ and approaches a limiting value of approximately $\sim 2r_{eff}$ from above. In contrast, the transverse blockade width exhibits a weaker dependence on optical depth and slowly approaches the same value from below. (see Appendix \ref{app:derivation}).
}
\label{fig:schrodinger solution}
\end{figure}

Figure~\ref{fig:schrodinger solution} shows the numerical solution of Eq.~\eqref{eq: schrodinger} in center-of-mass and relative coordinates, with the normalized wavefunction $\psi$ plotted for both the longitudinal and transverse directions. The depletion region at the center of Figs.~\ref{fig:schrodinger solution}(a,b) reveals the two different blockade and broadening mechanisms.
We again observe the interference effect already visible in Fig.~\ref{fig:two_photon_amplitude_trans}(b), where oscillations appear in the transverse direction, whereas the longitudinal plane shows only diffusion-like behavior with no oscillations [Fig.~\ref{fig:schrodinger solution}(c)].
Physically, the blockaded region acts as a thin wire (in our 2D geometry, with the relative coordinate $z$) that diffracts the light and redistributes its intensity in the output plane. This picture also makes the two-sided character of the transverse term explicit. Because the term is photonic and dissipation-free, the transverse blockade length is decoupled from the atomic response. It is fixed by the blockade radius and broadened only by vacuum diffraction, in contrast to the bandwidth-limited diffusive broadening in the longitudinal direction. The transverse correlation is therefore a direct probe of $\rb$.
 
To further confirm the scaling behavior, we scan the optical depth $\OD$ of the cloud, varied through the atomic density at fixed medium length $\len$, and extract the FWHM of the blockaded region in both directions [Fig.~\ref{fig:schrodinger solution}(d)]. Longitudinally, the width is diffusion-limited at low $\OD$ and approaches the blockade-limited floor at high $\OD$. The transverse width, in contrast, does not follow this atomic scaling but broadens through accumulated paraxial diffraction. Its $\OD$ dependence enters through the depth of the phase imprinted by the dispersive tail of $V^{(2)}$, while the propagation length and wavenumber enter through the subsequent free propagation of this phase, so that at the exit of the medium $w_{\perp}\propto(\OD\,\len/k)^{1/8}$, as given by Eq.~\eqref{eq:app-wlens}.
\section{Calculation of  Spatial and Temporal Correlation Function}

Having obtained the two-photon steady-state wavefunction, we now calculate the experimentally accessible correlation function $g^{(2)}(\tau,z_1,z_2)$. Here $\tau$ is the time delay between photons detected longitudinally, while the transverse direction carries no time correlation but rather a spatial correlation in the $(z_1,z_2)$-plane, obtained from the photon distribution recorded on a spatially resolving detector.
To compute it, we evolve the wavefunction after one photon is detected at the exit of the medium at $z_1$, and follow the time evolution of the second photon.

\begin{figure}[!t]
\includegraphics[width=\linewidth]{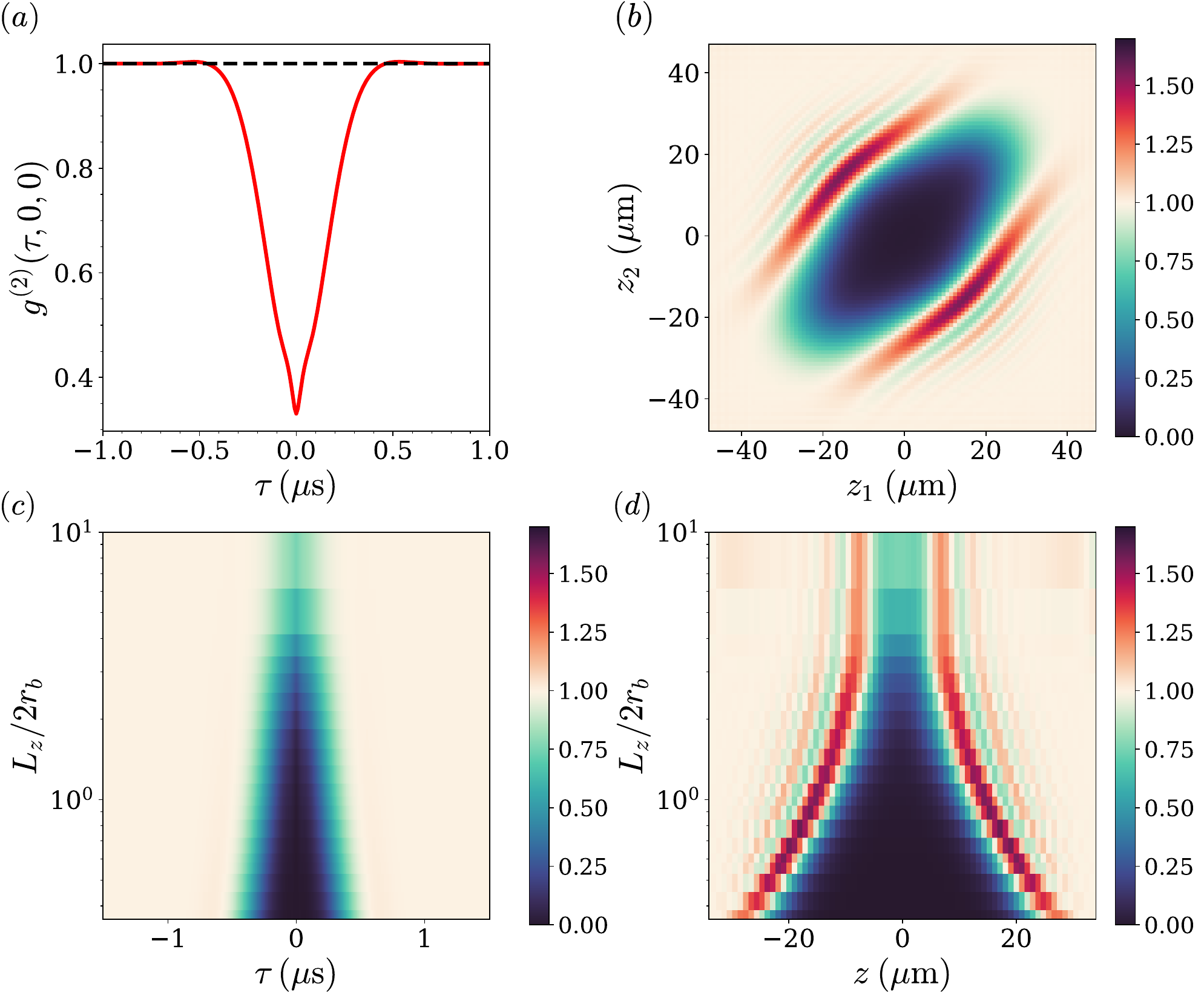}
\caption{Two-photon correlation function $g^{(2)}(\tau,z_1,z_2)$ of the wavefunction $\psi(x_1,z_1,x_2,z_2)$ in the temporal and spatial domains.
(a) $g^{(2)}(\tau,0,0)$ in the time domain at the center of the cloud, showing the longitudinal blockade. 
(b) $g^{(2)}(0,z_1,z_2)$ in the spatial domain, showing the transverse blockade.
(c,d) Dependence on the blockade radius relative to the transverse cloud width $L_z$, in the longitudinal (c) and transverse (d) directions, showing the reduction of the blockade interaction due to the transverse component.
For the transverse correlation, we take the diagonal line $z=z_1-z_2$. Simulation parameters are given in Sec .~\ref {numerical_params}.}
\label{fig:post_propgation}
\end{figure}

To compute these functions, we evolve the two-photon wavefunction [Eqs.~(\ref{eqs:full2})] over a time $\tau$ following the detection of the first photon at $x_1=\xo$, solving the equation of motion for the second photon [Eq.~\eqref{eq:one-photon}].
We denote the second-photon amplitudes as $e(\tau,z_1,x_2,z_2)$, $p(\tau,z_1,x_2,z_2)$, and $s(\tau,z_1,x_2,z_2)$, and integrate Eqs.~(\ref{eq:one-photon}) from $t=0$ to $t=\tau$, subject to the initial and boundary conditions
\begin{equation}
\begin{array}{lll}
e(0,z_1,x_2,z_2)&=& EE(x_1=\xo,z_1,x_2,z_2)\\
p(0,z_1,x_2,z_2)&=& EP(x_1=\xo,z_1,x_2,z_2)\\
s(0,z_1,x_2,z_2)&=& ES(x_1=\xo,z_1,x_2,z_2)\\
e(t,z_1,x_2=-\xo,z_2)&=& EE(x_1=\xo,z_1,x_2=-\xo,z_2)\:.
\end{array}
\end{equation}
After normalization, the solution gives $g^{(2)}(\tau,0,0)$ and $g^{(2)}(0,z_1,z_2)$, shown in Figs.~\ref{fig:post_propgation}(a) and (b).
The blockade-induced depletion of the wavefunction thus translates into correlations in the parameter space $(\tau,z_1-z_2)$, with anticorrelation appearing in both the temporal and spatial domains.
The temporal correlation width is broadened by the EIT bandwidth, whereas the spatial width is unaffected by diffusion. It is set mainly by the blockade radius and flanked by bunching fringes, the direct signature of the diffractive mechanism.

To probe the interplay between the one- and two-dimensional regimes, we vary the blockade radius relative to the transverse cloud width, keeping the transverse photon width larger than the cloud width so that the paraxial effect is negligible.
For $\rb>\sigma_z$, the entire cloud is blocked and no strong transverse blockade is possible. The system then behaves as a 1D system, with no transverse interaction between the photons.
For $\rb<\sigma_z$, in contrast, two Rydberg polaritons can interact in both the longitudinal and transverse directions. As the number of transversely interacting Rydberg excitations grows, more photons pass through the medium, reducing the longitudinal interaction,
since amplitude diffracted from neighboring transverse regions fills in the central channel even at $z_1=z_2=0$,
and thereby modifies the longitudinal blockade width and $g^{(2)}(0)$ [Figs.~\ref{fig:post_propgation}(c,d)].

\section{Numerical Parameters}\label{numerical_params}
\noindent We consider $^{87}$Rb atoms excited to the Rydberg level $70S_{1/2}$, with $C_6/\hbar=2\pi\times863$~GHz\,$\mu$m$^6$. All numerical calculations in this paper use the following parameters: $\sqrt{2\pi}\sigma_x=62.5$~$\mu$m, $\sqrt{2\pi}\sigma_z=31.25$~$\mu$m, $\rb=7.4$~$\mu$m, $\Omega=4$~MHz, $\Gamma=3.03$~MHz, $\gamma=0.07$~MHz, $\Delta=\delta=0$, and $\OD=60$, with $w_0=25$~$\mu$m for Figs.~\ref{fig:two_photon_amplitude} and~\ref{fig:two_photon_amplitude_trans}.
All frequencies are in the half-width convention. In Fig.~\ref{w0 variation}, $w_0$ is varied from $\rb/5$ to $5\rb$, and in Figs.~\ref{fig:post_propgation}(c,d), $\rb$ is varied from $1.2$~$\mu$m to $44.28$~$\mu$m.

\section{Discussion and Outlook}

We have developed a theory of photon-photon correlations in a Rydberg medium beyond the one-dimensional approximation, in the regime where the beam diameter exceeds the blockade radius.
The full two-photon wavefunction, computed for experimentally realistic parameters, shows that the blockade imprints correlations in the transverse spatial domain alongside the familiar temporal ones.
Reducing the nine-component model to the single diffusive-diffractive Schr\"odinger equation [Eq.~\eqref{eq: schrodinger}] identifies the correlation mechanism in each direction. Along the propagation axis it is atomic, bandwidth-limited diffusion, and in the transverse direction imaginary photonic diffraction. 

Our study shows the flexibility to tune the transverse Rydberg interaction by controlling the beam waist, interaction strength, and atomic density distributions. Furthermore, this study establishes that the transverse correlation can serve as a new measurable observable beyond traditional temporal correlations and can also be used as a direct method to measure the blockade radius in the dissipative case.  

Several directions follow naturally.
A full 3D treatment, retaining both transverse dimensions, would resolve the azimuthal structure of the correlations and test the radial symmetry of the transverse blockade.
Moving from the dissipative to the dispersive regime opens the prospect of topological defects in the spatial domain, complementing the recently observed temporal quantum vortices \cite{drori2023quantum,das2025multiband}.
Together with spatially resolved single-photon detection, these developments position the transverse Rydberg blockade as a resource for generating and probing higher-dimensional quantum states of light.
\section{Acknowledgements}
D.S, M.M, W.A, S.H, acknowledge the support by the European Union’s Horizon 2020 program under the ERC grant SUPERWAVE (grant No.101071882) and by the Deutsche Forschungsgemeinschaft (DFG) within the collaborative research center SFB/TR185 OSCAR, project A8 (No. 277625399). B.C.D and O.F acknowledge financial support from the Israel Science Foundation (grant No.~3491/21, 1982/22, 3534/26), the Minerva Foundation with funding from the Federal German Ministry for Education and Research, the Leona M.~and Harry B.~Helmsley Charitable Trust, the Shimon and Golde Picker - Weizmann Annual Grant, and the Laboratory in Memory of Leon and Blacky Broder.

\bibliography{bibliography}
\newpage\clearpage

\onecolumngrid 
\setcounter{figure}{0}
\renewcommand{\thefigure}{S\arabic{figure}}
\setcounter{equation}{0}
\renewcommand{\theequation}{S\arabic{equation}}

\appendix
\section{Appendix: Width of the depletion region}
\label{sec:app-width}\label{app:derivation}
The effective interaction $V^{(2)}$ separates into a dissipative and a dispersive part,
\begin{equation}
V^{(2)}(r)=\frac{1}{1-iu^{6}}
=\frac{1}{1+u^{12}}
+\,i\,\frac{u^{6}}{1+u^{12}},
\qquad u=\frac{r}{\tilde{\rb}},
\label{eq:app-Vsplit}
\end{equation}
where $r=\sqrt{x^{2}+z^{2}}$. The real part is absorptive and falls off as $u^{-12}$, whereas the dispersive imaginary part falls off far more slowly, as $u^{-6}$. This asymmetry is what makes the two directions behave differently.

Since $V^{(2)}$ depends only on $r$, it is isotropic by itself and imprints the same depletion in every direction. Where the kinetic terms of Eq.~\eqref{eq: schrodinger} are negligible, the equation integrates exactly to $\psi=\exp[-(\OD\,X/\len)\,V^{(2)}]$. Setting $|\psi|^{2}$ to half its background value gives the common width
\begin{equation}
w_{\rm common}=2\,r_{\rm eff}(X),
\qquad
r_{\rm eff}(X)=\tilde{\rb}\left[\frac{2\,\OD\,X}{\len\ln 2}-1\right]^{1/12},
\label{eq:app-common}
\end{equation}
which is simply the blockade core dressed by the absorptive part of $V^{(2)}$ and carries no directional information. The anisotropy enters only through the two kinetic terms of Eq.~\eqref{eq: schrodinger}, which modify $w_{\rm common}$ differently in the two directions.

\subsection{Transverse blockade width}
\label{app:lensing}

The imaginary part of $V^{(2)}$ imprints a phase $\varphi(X,z)=-(\OD\,X/\len)(\tilde{\rb}/z)^{6}$, which the transverse kinetic term converts into a redistribution of intensity. Writing $\psi=\sqrt{I}\,e^{i\varphi}$, the paraxial term of Eq.~\eqref{eq: schrodinger} yields the transport equation
\begin{equation}
\partial_{X}I=-\frac{2}{k}\,\partial_{z}\!\left(I\,\partial_{z}\varphi\right),
\label{eq:app-transport}
\end{equation}
so the dispersive tail redistributes intensity without absorbing it. Evaluating Eq.~\eqref{eq:app-transport} in the undepleted tail, $I\simeq1$, with $\partial_{z}^{2}z^{-6}=42\,z^{-8}$ and $\varphi$ linear in $X$, gives an accumulated redistribution $\delta I=42\,\OD\,X^{2}\tilde{\rb}^{6}/k\len z^{8}$. Considering $\delta I=1/2$ then gives the modified blockade width
\begin{equation}
w_{\perp}=2\left[\,84\,\frac{\OD\,X^{2}\tilde{\rb}^{6}}{k\,\len}\,\right]^{1/8}
\;\propto\;\OD^{1/8}X^{1/4}.
\label{eq:app-wlens}
\end{equation}
The diffractive term therefore broadens $w_{\rm common}$ only weakly, so the transverse correlation length remains set by the blockade radius, dressed by the interaction tail, and decoupled from the EIT bandwidth.

The same lensing also shapes the profile around the dip. Since $\partial_{z}^{2}\varphi<0$ in the tail, the intensity removed from the blockaded core accumulates on either side of it, which is the bunching flanking the transverse dip in Fig.~\ref{fig:post_propgation}(b). On top of this, the diffracted and undiffracted parts of the field interfere, producing the oscillations visible in Figs.~\ref{fig:two_photon_amplitude_trans}(b) and~\ref{fig:schrodinger solution}(b) and carried through to the correlation function.

\subsection{Longitudinal blockade width}
\label{app:longitudinal}
In the longitudinal direction the second-derivative coefficient of Eq.~\eqref{eq: schrodinger} is real, $D=4\len/\OD$, so the depletion created by the blockade spreads diffusively,
\begin{equation}
w_{\parallel}(X)\simeq\sqrt{2DX}=\sqrt{\frac{8\len X}{\OD}},
\label{eq:app_wdiff}
\end{equation}
which at the cloud exit ($X=\len$) gives the $1/\sqrt{\OD}$ scaling of Fig.~\ref{fig:schrodinger solution}(d).
Unlike the paraxial term, a real diffusion kernel damps phase gradients rather than converting them into intensity, so there is no longitudinal analog of the lensing mechanism. The only competing scale is the dissipative floor of Eq.~\eqref{eq:app-common}, which grows slowly with $\OD$.
The longitudinal width is therefore diffusion-dominated, $\propto1/\sqrt{\OD}$, at moderate optical depth, and saturates toward the interaction-tail floor once the two contributions become comparable, at $\OD\sim10^{2}$ for our parameters, consistent with the flattening of the longitudinal curve in Fig.~\ref{fig:schrodinger solution}(d).
The opposite trends in the two widths with $\OD$, longitudinal shrinking and transverse slow growth, thus share a single origin. The coefficient of the relative-coordinate kinetic term is real (atomic, bandwidth-limited) along the propagation direction and imaginary (photonic, dissipation-free) transverse to it.
\end{document}